\documentclass[12pt]{article}
\usepackage{epsfig}
\begin{document}
\def\be{\begin{equation}}
\def\ee{\end{equation}}
\def\bea{\begin{eqnarray}}
\def\eea{\end{eqnarray}}
\def\bear{\be\begin{array}}
\newcommand{\gtrsim}{\stackrel{>}{\sim}}
\newcommand{\lessim}{\stackrel{<}{\sim}}
\newcommand{\dst}{\displaystyle}
\newcommand{\bm}{\boldmath}
\newcommand{\fr}[2]{\frac{{\dst #1}}{{\dst #2}}}
\newcommand{\fn}[1]{\footnote{{\normalsize #1}}}
\newcommand{\vep}{\varepsilon}
\newcommand{\epe}{\mbox{$e^+e^-\,$}}
\newcommand{\ggam}{\mbox{$\gamma\gamma\,$}}
\newcommand{\egam}{\mbox{$\gamma e\,$}}
\newcommand{\eeww}{\mbox{$e^+e^-\to W^+W^-\,$}}
\newcommand{\ggww}{\mbox{$\gamma\gamma\to W^+W^-\,$}}
\newcommand{\gewnu}{\mbox{$e\gamma\to \nu W\,$}}
\newcommand{\ggzz}{\mbox{$\gamma\gamma\to ZZ\,$}}
\newcommand{\geeww}{\mbox{$\gamma e\to  W^+W^- e\,$}}
\newcommand{\egeh}{\mbox{$e\gamma\to eH\,$}}
\newcommand{\SM}{${\cal S} {\cal M}\;$}
\newcommand{\DSM}{$2{\cal H} {\cal D} {\cal M} \;$}
\newcommand{\MSM}{${\cal M} {\cal S}  {\cal M}\;$}
\newcommand{\CP}{${\cal C}{\cal P}\;$}
\newcommand{\MSSM}{${\cal M}{\cal S} {\cal S}  {\cal M}\;$}
\newcommand{\SU}{${\cal S}{\cal U} 2\otimes{\cal U} 1\;$}
\date{}

\title{{\bf Why Photon Colliders are necessary
in a future collider program}\\ {\em\normalsize to be published in
Proc. GG2000, DESY, June 2000, Nucl. Instr. Meth.}}

\author{Ilya F. Ginzburg\\
Sobolev Institute of Mathematics SB RAS, 630090, Novosibirsk,
Russia}

\maketitle

\centerline{\bf Subjects of discussion}
\begin{enumerate}
\item Different scenarios
\vspace{-0.3cm}
\item Higgs window to a New Physics
\vspace{-0.3cm}
\item Anomalies in the interactions of gauge and Higgs bosons
\vspace{-0.3cm}
\item Some problems with $t$--quarks and \MSSM.
\vspace{-0.3cm}
\item Some problems in QCD and hadron physics.
\vspace{-0.3cm}
\item By-product: Production of axions, etc. from region of
  conversion $e\to\gamma$.
\end{enumerate}

\section{Different scenarios}

{\bf Photon Colliders in the widespread scenario.}
When discussing the program for future high energy colliders, the
basic point is usually that {\em Nature is so favorable to us
to dispose the essential fraction of new particles and the
thresholds of new interactions} (e.g., effects of new higher
dimensions, compositeness, etc.) {\em within the LHC operation
domain.}

   In this case main discoveries will be made at the
Tevatron and the LHC. The $e^+e^-$ Linear Colliders (LC) in their
first stages will be machines for measuring precise values of
coupling constants and exploring in detail supersymmetry. The high
luminosity expected for \epe Linear Colliders provides opportunity
to obtain parameters of models realized with very high accuracy,
see, e.g., \cite{Zer}.

   With the Photon Collider mode of LC \cite{GKST} having roughly the
same luminosity as that for \epe\ mode \cite{Final} these results
will be improved. Indeed,

$\bullet$ The cross sections of production of pairs of charged
particles in the \ggam\ collision are $5\div 8$ times higher than
that in the \epe\ collision in its maxima; this ratio increases
with growth of energy.\\ $\bullet$ The set of final states at a
Photon Collider is much richer than that at in an \epe\ mode.\\
$\bullet$ One can vary polarizations of photon beams relatively
easily ({\em circular $\to$ transverse}). \\

Unfortunately, this Photon Collider opportunity was not explored
by the community in necessary detail. It can be assured that the
forthcoming studies of many physicists with detailed simulation
will show an exceptional potential for the Photon Collider mode in
both new problems and all problems considered till now.

{\bf\bm Possible \SM -- like  scenario.} It can also happen that
the opposite scenario will be realized:\\ {\bf\bm No new particles
and interactions will be discovered at the Tevatron, LHC and \epe\
LC, except the Higgs boson.} If additionally its coupling constant
to quarks and gauge bosons will be close to their values in the
\SM, we find a \SM-- like picture of the World.  It can be
realized both if our World is really described by simple \SM up to
very small distances and if some other model is realized and New
Physics is {\em round the corner}. In this case the main goal of
studies at new colliders will be the hunting for indirect signals
of New Physics -- deviations of observed quantities from \SM
predictions.\\ {\it Photon Colliders are the best machines for the
hunting for signals of New Physics if the \SM-- like scenario is
realized.}

\section{Higgs window to a New Physics}

The study of Higgs-boson couplings with photons ($h\gamma\gamma$
and $hZ\gamma$) looks like a very promising tool for resolving of
models of New Physics.

$\bullet$ These couplings are absent in the \SM at tree level,
appearing only at the loop level. Therefore, the background for
signals of New Physics will be relatively lower here than in other
processes which are allowed at tree level of the SM.
\\ $\bullet$ All fundamental charged particles contribute to these
effective couplings. The whole structure of the theory influences
the corresponding Higgs-boson decays.
\\ $\bullet$ The expected accuracy in the two-photon width is
{\em about 2\%} at  $M_h\le 150$ GeV and even at the luminosity
integral $\sim 30$ fb$^{-1}$ in the high energy peak of the luminosity
spectrum \cite{Jik}.

In the \DSM and \MSSM, the observed Higgs boson will be either the
lightest Higgs boson $h$ or heavier one $H$. Assuming the coupling
constants of observed Higgs boson to quarks and gauge bosons are
close to their values in the \SM, other Higgs bosons can easily
avoid observation at LHC and \epe\ LC due to small couplings to
standard matter. These neutral Higgses can be seen in the process
$\ggam\to h$ \cite{spir} and charged Higgses in the process
$\ggam\to H^+H^-$.

Even if these additional Higgs bosons are so heavy that they
cannot be produced at a first stage  Photon Collider, the
models can be distinguished well via precise measurement of
two--photon width of observed \SM--like Higgs boson \cite{GKO}. In
this paper we assumed that the discussed \SM--like scenario is
realized with an accuracy estimated for \SM Higgs boson at an \epe\
Linear Collider \cite{Zer}. The two photon width is calculated via
the Higgs couplings with the matter measured at an \epe\  LC. In the
\DSM, deviation from \SM is due mainly to  contribution of
charged Higgs. In the general case the two-photon width is about
10\% less than that in the \SM (see Figs. 1). It is several times
larger than the expected experimental inaccuracy. Therefore,
measurement of the two-photon width at a Photon collider can resolve
these models reliably.

\begin{figure}[hbt]
\epsfig{file=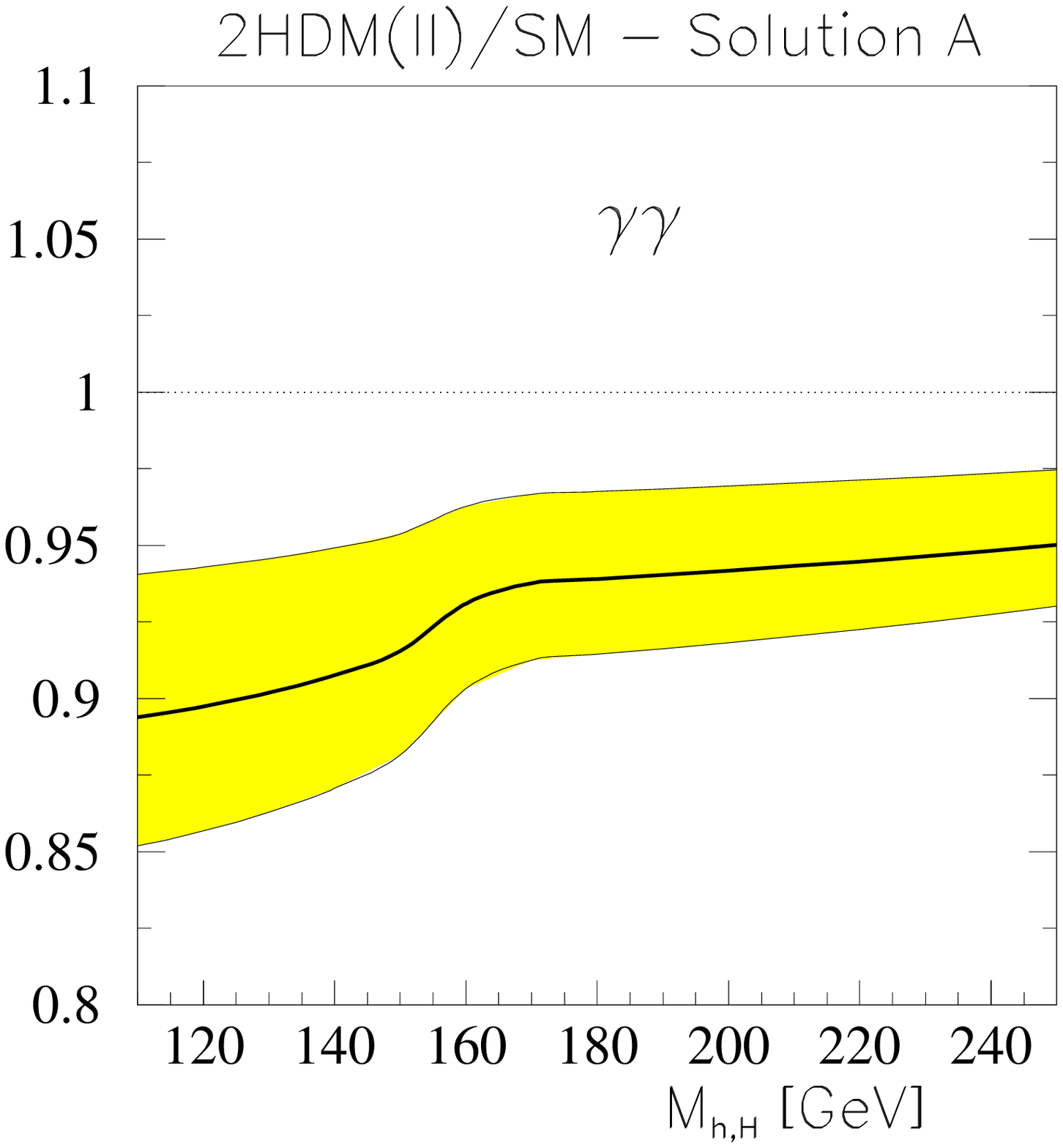,width=7.5cm}
\epsfig{file=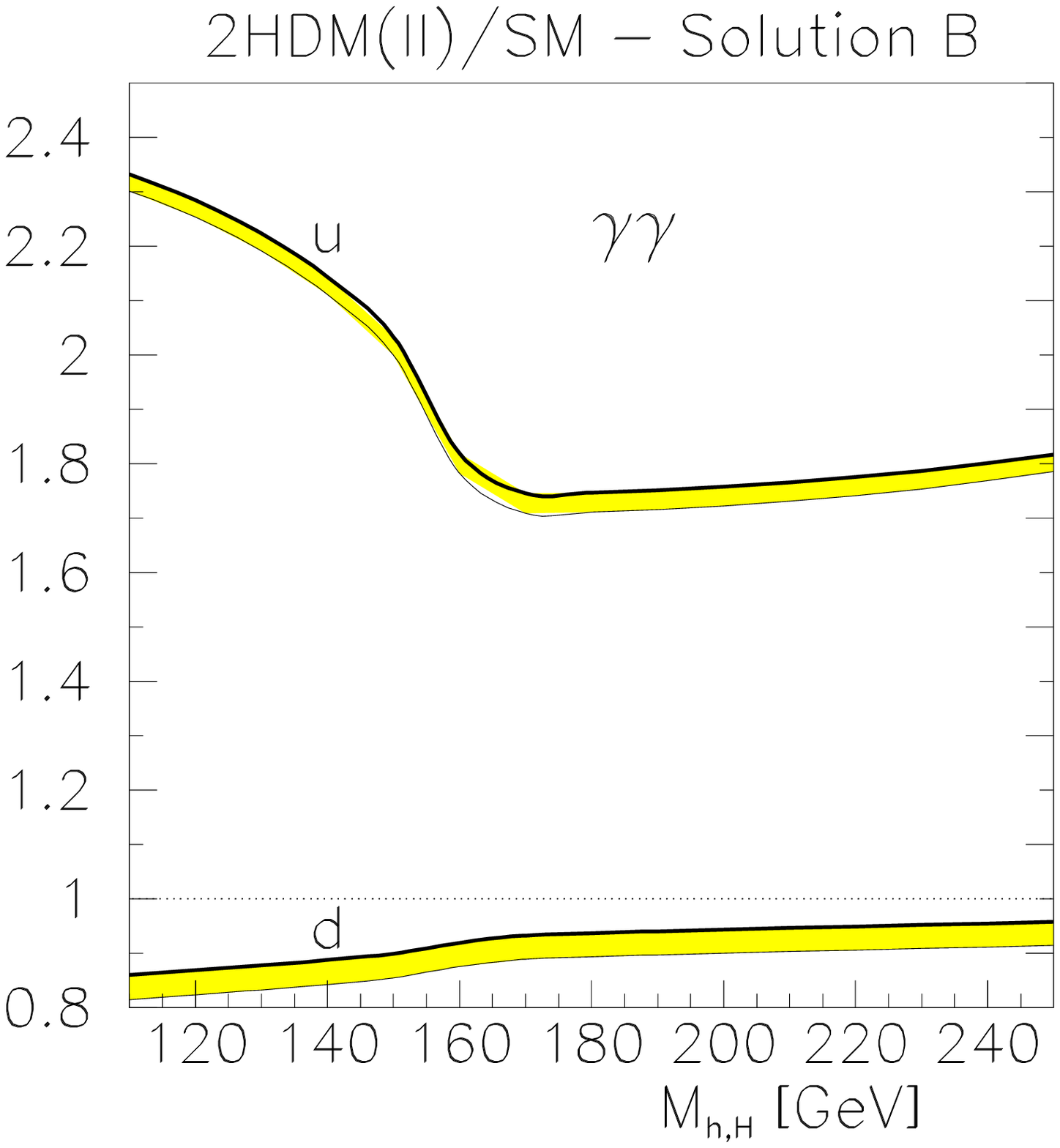,width=7.5cm}
\caption{\em The ratio of two--photon width of Higgs boson in \DSM
to its \SM value at $\lambda_5=0$ and $M_h\ge 800$ GeV. Shaded
zones correspond to experimental uncertainties expected in the
experiments at \epe\  LC. Deviation from \SM depends on
$\lambda_5$ as $\propto(1-\lambda_5v^2/M_{H^\pm}^2$) with $v=246$
GeV is v.e.v. of Higgs field. Solution A is really close to \SM.
For the solution B some of the couplings of the Higgs field with matter are
close to the \SM values, the others have the opposite sign. }
\end{figure}

In the \MSSM the lightest Higgs boson is decoupled with
superpartners and other members of Higgs multiplet if they are
very heavy. More detailed calculations will give us the upper bounds
for masses of superparticles which can influence the photon widths
so that the difference will be seen in the experiment. Preliminary
estimates show that these values are higher than the discovery
limits at LHC.

In many variants of \DSM or \MSSM the masses of heavy scalar
Higgs $H$ and Higgs pseudoscalar $A$ are close each other. In
some other variants they are mixed (\CP violated scenarios). The
observations at LHC and \epe\ Linear Collider cannot often
resolve these opportunities due to low resolution for these
bosons. The polarization asymmetries in Higgs boson production
at a Photon Collider can resolve these variants, i.e.  establish
whether \CP parity at Higgs level is violated or not.

\section{Anomalies in the interactions of gauge and Higgs bosons}

Before discovery new heavy particles inherent New Physics, it
reveals itself at lower energies as some  anomalies in the
interactions of known particles. Our goal is to find these
anomalies and discriminate tham as best as possible. {\em The
correlation between coefficients of different anomalies will be
the key for understanding what is the nature of New Physics.}

{\bf Interactions of gauge bosons.}
The practically unique process under study in the \epe mode is
\eeww. At suitable electron polarization the neutrino exchange
contribution (having small interest) disappears, and the residual cross
section (obliged by photon and $Z$ boson exchange) has a maximum
about 2 pb within the LEP operation interval and decreases with energy
after that. The cross sections of other processes with $W$
production ($\epe\to \epe WW$, $\epe\to e\nu W$, ...) are small at
$\sqrt{s}<1$ TeV.

At Photon collider the main processes are -- \ggww, \gewnu.
Their cross sections are about 80 pb and are energy independent at
$\sqrt{s}>200$ GeV. That is at least 40 times higher than for
\epe\ collisions, it gives about $(1\div 5)\cdot10^7$ $W$'s per
year, comparable with $Z$ production at LEP. Due to high
value of these basic cross sections, many processes of 3-rd and
4-th order have large enough cross sections:\\ {\bm $\egam\to
eWW$, $\ggam\to ZWW$, $\egam\to \nu WZ$, $\ggam\to WWWW$,
$\ggam\to WWZZ$, etc.}
Large variety of these processes permit us to discover and separate
well anomalies in specific processes and (or) distributions. This
subject needs studies of all processes enumerated with details
of behavior in different regions of phase space and polarization
dependence.
\begin{figure}[hbt]
\centering \vspace*{-0.0cm}
\epsfig{file=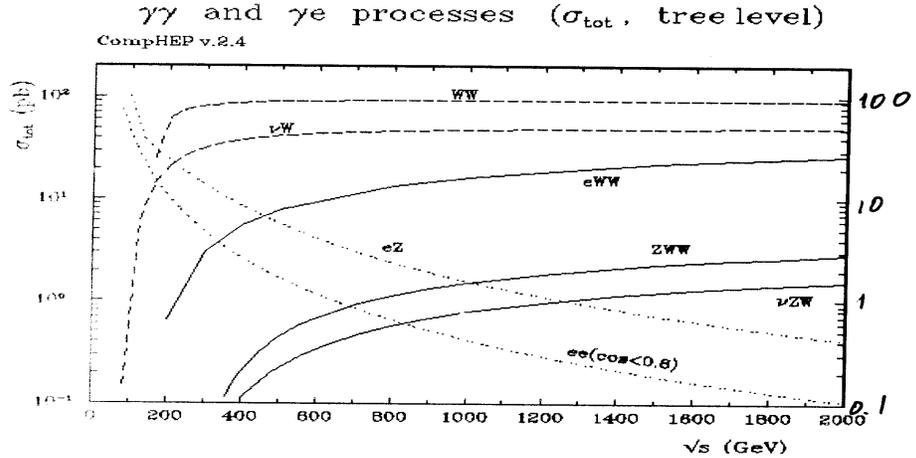,width=12cm,height=6cm} \caption{\em Some
processes of  gauge boson production at \egam\ and \ggam\
colliders}
\end{figure}

$\bullet$ For example, the following line of procedures is almost
evident:\\ (a) To extract $\gamma WW$ anomalies from $\egam\to \nu
W$.\\ (b) To extract $ZWW$ anomalies from \eeww.\\ (c) To extract
$\ggam WW$ anomaly from \ggww.\\
$\bullet$ {\bm \gewnu}. The cross section of process is $\propto
(1-2\lambda_e)$, it is switched on or off with variation of
electron helicity $\lambda_e$. It gives precise test of absence of
right handed currents in the interaction of $W$ with the matter.\\
$\bullet$ {\bf\bm $\egam\to eWW$.} The cross section of process at
$\sqrt{s}=500$ GeV is about 10 pb, corresponding counting rate is
few millions events per year.\\$\star$ In events with transverse
momentum of scattered electron $p_\bot\ge 30$ GeV one can study
here anomaly $\gamma ZWW$.\\ $\star$ The charge asymmetry of
produced $W$'s looks like most sensitive key for the study of strong
interaction in Higgs sector.\\
$\bullet$ {\bm\bf \ggww and \gewnu}. The two--loop radiative
corrections to these processes  are measurable and sensitive to
the problems:\\ $\star$ {\em construction of $S$--matrix of theory
with unstable particles;}\\ $\star$ {\em gluon corrections like
Pomeron exchange between quark components of $W$'s.}

{\bf\bm Interactions of Higgs boson with light ( $\ggam\to h$, $\egam\to
eh$).}
All observable anomalous \CP-- even and \CP-- odd interactions of
Higgs boson with light are summarized in the form of an effective
Lagrangian:
\bear{c}
 \Delta {\cal L} = 2Hv\left(\theta_\gamma\fr{F_{\mu
\nu } F^{\mu\nu}} {2\Lambda_{\gamma}^2} + \theta_Z \fr{Z_{\mu \nu
} F^{\mu \nu}}{\Lambda_Z^2}+\right. \\
 \left.
i\theta_{P\gamma}\fr{F_{\mu \nu } \tilde{F}^{\mu\nu}}
{2\Lambda_{P\gamma}^2} + i\theta_{PZ}
 \fr{Z_{\mu \nu } \tilde{F}^{\mu \nu}} {\Lambda_{PZ}^2}\right),
\;\; \left(\theta_i=e^{i\xi_i}\right).
\end{array}
\ee
Here $F^{\mu\nu}$ and $Z^{\mu\nu}$ are the standard field
strengths for the electromagnetic and $Z$ field,
$\tilde{F}^{\mu\nu} =\varepsilon^{\mu\nu\alpha\beta}
F_{\alpha\beta}/2$ and {\em $\xi_i$ are the phases of couplings},
generally different from 0 or $\pi$ even in the C--even case (due
to possible anomalous contributions of some light particles)
\cite{GIIv}.

While \CP even anomalies can be seen via the values of
measured cross sections, the \CP anomalies will be seen via the
polarization asymmetries ("longitudinal" -- variation of cross
sections with change of sign of collided photon helicities, or
"transverse" -- variation of cross section in dependence on the
angle between directions of linear polarization of collided
 photons). The effects are large enough to see them at
reasonable values of anomaly scales $\Lambda_i$ and phases
$\xi_i$ (see Fig. 2).

\begin{figure}[hbt]
\centering \vspace*{-0.0cm} \epsfig{file=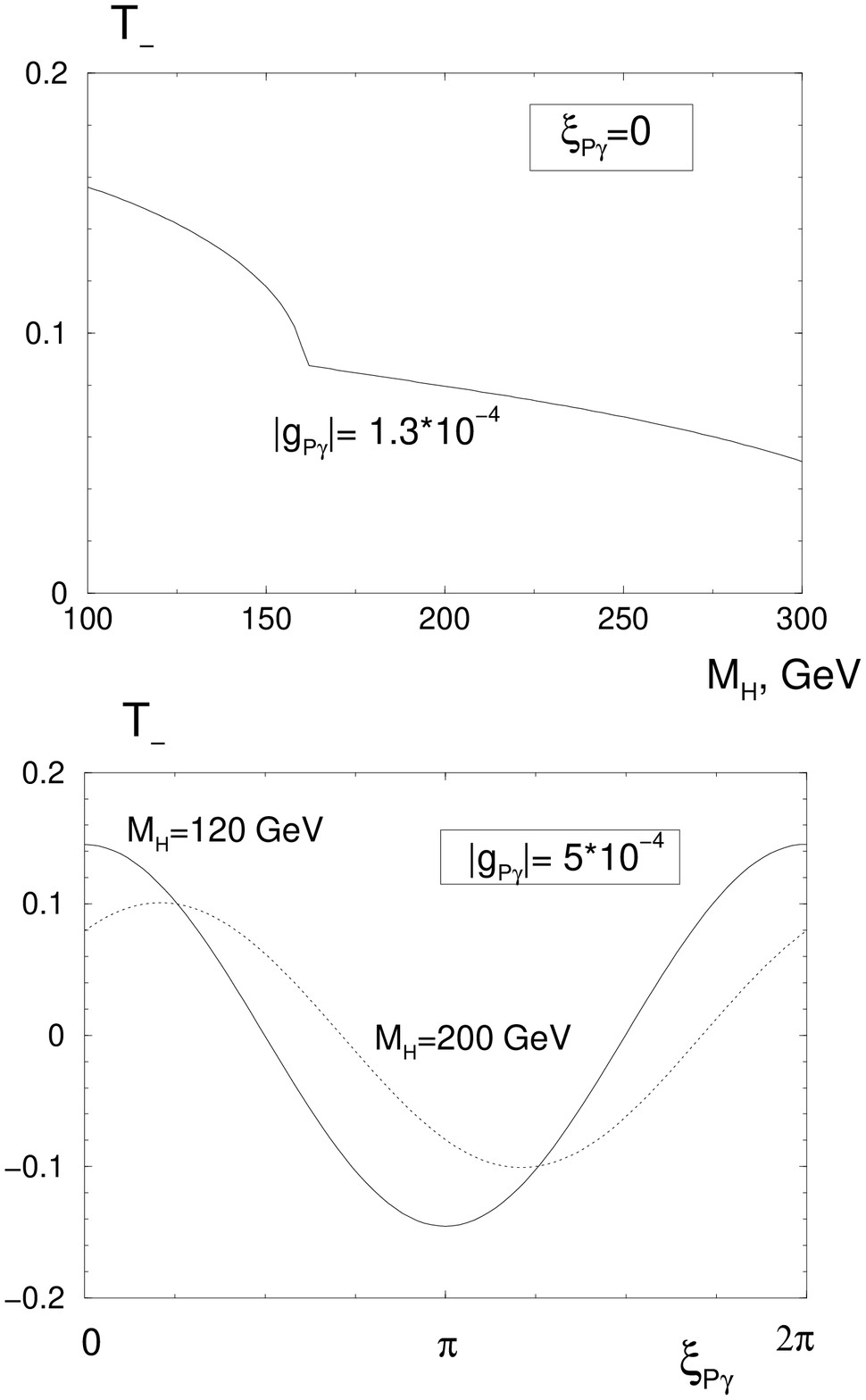,height=12cm}
\caption{\em "Longitudinal" asymmetry for $\ggam\to H$ process}
\end{figure}

The specific example of anomaly presents \CP violating mixing
in Higgs sector of \DSM (scalar -- pseudoscalar) with mixing angle
$\alpha_2$. We present results in Fig. 3 fixing other model parameters to be
close to the \SM case.

\begin{figure}[hbt]
\centering \vspace*{-0.0cm}
\epsfig{file=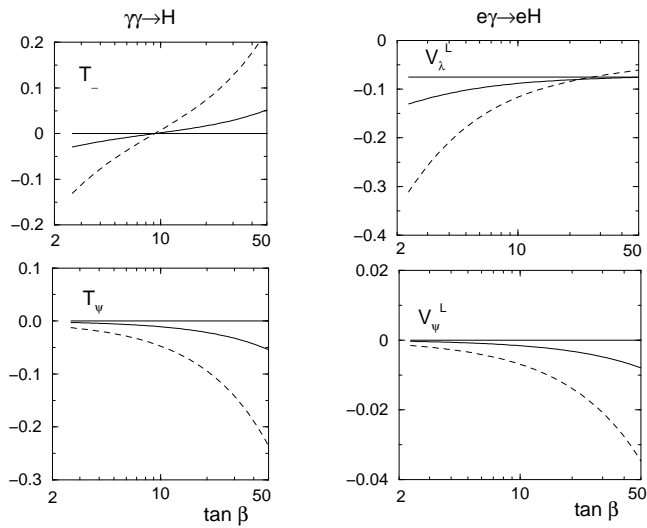,height=7cm,angle=0}
 \caption {\em Polarization asymmetries in $\ggam\to h$ and
$\egam\to eh$ processes due to scalar-pseudoscalar mixing in
\DSM(II), $M_h=110$ GeV; for \egeh process $E_{tot}=1.5$ TeV.
The thick solid lines show the \SM values; thin solid and dashed
lines refer to $\sin\alpha_2 = 0.1$ and $0.5$ respectively
($\alpha_2$ is light scalar--pseudoscalar mixing angle).}
\end{figure}

$\bullet$ It can happen that the masses of heavy scalar Higgs $H$
and pseudoscalar $A$ are so close to each other that they cannot be
resolved in the mass spectrum. The study of decay products cannot
distinguish this overlapping from the case of their mixing with
\CP violation. In the case when these Higgses  overlap in the
true mass spectrum, the correlations in the decay products will be
identical if \CP violated mixing exists or not. The study of
polarization asymmetries in the Higgs boson production at Photon
Colliders distinguish these two opportunities well.

\section{\bm $t$ -- quarks, \MSSM}

Here I present only few examples in which Photon Colliders look
absolutely necessary.

$\bullet$ {\bf\bm $t$--quarks}. In addition to the usually
discussed problems related to $t$-quarks, the specific one is the
study of axial anomaly in the process $\egam\to et\bar{t}$. It was
found that at small transverse momenta of scattered electrons
$p_\bot$ the cross section of subprocess $Z_L\gamma\to t\bar{t}$
with longitudinally polarized $Z$ does not disappear as  happens with
photons but diverges as $M_t^2/p_\bot^2$ \cite{GIl}\\

$\bullet$ {\bf\bm Some problems with \MSSM.}\\
$\star$ Due to high values of the basic cross sections for the
production of pairs of charged particles, Photon Colliders promise
to be an excellent place for observation of (even small) possible
flavor nonconservation in the neutral currents with
superparticles.\\
$\star$ If the stop squark is not very heavy, the atom-like
stoponium with mass 200-400 GeV should also exists. It will be
very narrow and hard to observe at hadron collider and \epe\
Linear Colliders. However, it can be clearly seen at \ggam collider
with cross section averaged over photon spectrum {\bf\bm
$<\sigma>\approx 10-50$ fb} and clear enough signature
\cite{GorIl}.\\
$\star$ Gaugino discovery. The expected discovery limit for
gaugino at LHC is about 170 GeV in the variants of MSSM which
looks most probable now. It can be enhanced till 450 GeV in the
"less probable" variants of MSSM.

\section{ QCD and Hadron Physics}

All problems studied at HERA and LEP will be studied at Photon Colliers but in
much more wide interval of parameters and with much better
accuracy. Among them I underline those which look most interesting
now.

$\bullet$ Study of photon structure function(s) at small $x$.\\
$\bullet$ Nature of growth of {\bf total cross sections}. The
widespread concepts assume standard Regge type factorization and
universal energy behavior for different processes. With Photon
Colliders --{\em for the first time in particle physics}-- one can
have the set of mass shell cross sections of very high energy
processes, appropriate for the testing of factorization or the level
of its violation. They are $\sigma_{pp}$, measured at Tevatron and
LHC,  $\sigma_{\gamma p}$, measured at HERA,
$\sigma_{\gamma\gamma}$, measurable at Photon Collider. For this
goal, {\em the preliminary stage of operations with low luminosity
can be used} to observe large enough cross sections at small
scattering angles.\\
$\bullet$ In this very low luminosity stage the study of events
with particle production in the center of rapidity scale and with
rapidity gaps for two vector mesons or dijets originated from
initial photons will be crucial for the understanding of {\bf
double Pomeron effects} having no reliable explanation till now.\\
$\bullet$ The study of {\bf charge asymmetry of produced hadrons}
in \ggam\ collisions will give quite new information about
quark--gluon matter at small distances. The charge asymmetry of
the produced hadrons in the \egam\ collisions with transverse
momentum of scattered electron $p_\bot\ge 30$ GeV will show in
explicit form the relation between hadron states produced by
vector and axial current.

\section{Using of conversion region as \egam collider}

The conversion region in \egam\  collider with c.m.s. energy about
1.2 MeV but with luminosity about 0.1 fb$^{-1}$/sec! It will
be unique source of light goldstone particles (axions,
majorons, etc.) $LGP$, expected in numerous schemes \cite{Pol}.\\
~~~~The production processes are
\be
e\gamma_0\to e(LGP)\,,\;\;\;\; \gamma\gamma_0\to (LGP)\,.
\ee
The energy of $LGP$ is limited from above as
\be
E_{LGP}\le \fr{x+a^2+\sqrt{(x-a^2)^2-4a^2}}{2(x+1)}E,\quad
a=\fr{m_a}{m_e}.
\ee
The angular spread of these  $LGP$'s is even narrower than that of
high energy photons. In comparison with neutrinos produced by photons
in the final wall, the main part of these "axions" is concentrated
in a $10^6\div 10^8$ times more narrow solid angle.
\begin{figure}[hbt]
 \centering \vspace*{-0.0cm}
\epsfig{file=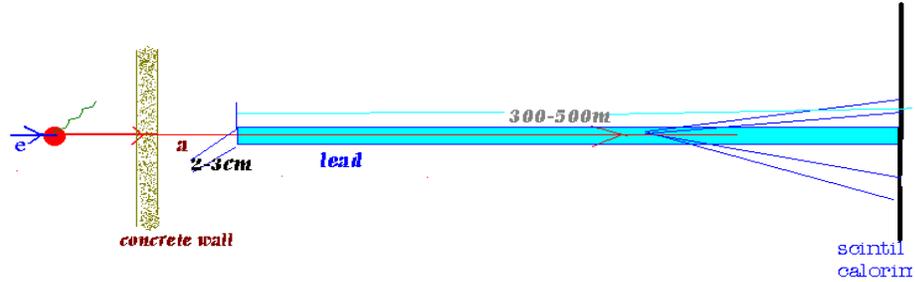,height=4cm,width=12cm}
\caption{Possible scheme for recording of the "axion"}
\end{figure}

These {\em $LGP$}s interact with  matter very weakly.  So, the
registration scheme can be of the type shown in Fig. 4 (where
magnetic bending of residual electrons after conversion is also
shown): the $LGP$  produced after concrete wall travel
to the detector inside  $L=300-500$ m long
lead wire of diameter 3-5 cm. Here it can interact with
hadrons in the nucleus of lead and produce hadrons with typical mean
transverse momentum about 300-500 MeV. They can be detected in
scintillator-like devices of diameter 3-5 m in the end of wire.

\section{Conclusion}

This discussion shows that {\em R\&D for Photon collider
mode should be performed simultaneously with that of \epe mode of
TESLA, etc. Final decision about the turn of different stages of
Linear collider should be made only {\bf AFTER} first
operations of LHC.} It can imagine even the opportunity that the
Photon collider mode will be switched on before \epe\ mode. The
advantages are:\\ $\bullet$ The basic electron energy
is lower.\\ $\bullet$ Positron beam is unnecessary.

I am thankful  V.Ilyin, I. Ivanov, M.Krawczyk, P.Olsen for
collaboration related different parts of paper and A.Djouadi, V.
Serbo, M. Spira, V. Telnov, P.Zerwas for useful discussions. I
am grateful to RFBR (grants 99-02-17211 and 00-15-96691) for
support and DESY direktorat for the sponsoring my participation
in the Workshop.

\end{document}